\documentclass{ws-ijmpa}

\begin{document}

\renewcommand{\draftnote}{} 
\renewcommand{\trimmarks}{} 

\markboth{E. A. Matute} {Twin and Mirror Symmetries from
Presymmetry}

\catchline{}{}{}{}{}

\title{TWIN AND MIRROR SYMMETRIES FROM PRESYMMETRY}

\author{\footnotesize ERNESTO A. MATUTE}

\address{Departamento de F\'{\i}sica, Universidad de Santiago de
Chile,\\ Usach, Casilla 307 -- Correo 2, Santiago, Chile\\
ernesto.matute@usach.cl\\[11pt]}

\maketitle

\pub{}{}

\begin{abstract}
We argue that presymmetry, a hidden predynamical electroweak
quark--lepton symmetry that explains the fractional charges and
triplication of families, must be extended beyond the Standard
Model as to have a residual presymmetry that embraces partner
particles and includes the strong sector, so accounting for the
twin or mirror partners proposed to alleviate the naturalness
problem of the weak scale. It leads to the full duplication of
fermions and gauge bosons of the Standard Model independently of
the ultraviolet completion of the theory, even if the Higgs
particle is discarded by experiment, which adds robustness to twin
and mirror symmetries. The established connection is so strongly
motivated that the search for twin or mirror matter becomes the
possible test of presymmetry. If the physics beyond the Standard
Model repairs its left--right asymmetry, mirror symmetry should be
the one realized in nature.

\end{abstract}

\ccode{PACS numbers: 11.30.Hv, 11.30.Ly, 12.60.Fr, 12.60.Cn}

\section{Introduction}

The Standard Model (SM) of strong and electroweak interactions
with gauge symmetry $\mbox{SU}(3)_c \times \mbox{SU}(2)_L \times
\mbox{U}(1)_Y$ and extended with right-handed neutrinos has a
remarkable success in explaining all experimental data obtained so
far. To be completely accepted, however, the predicted neutral
Higgs boson has to be discovered. Electroweak high precision
measurements imply a light Higgs well below 1 TeV and a lower
bound on the scale of any new physics of several TeV.\cite{PDG}
If the SM is viewed as an effective theory with a cutoff about
this scale, a confrontation with the naturalness problem of the
weak scale is unavoidable, due to the quadratic divergences on the
cutoff that affect the Higgs mass.\cite{Barbieri} Clearly, the
new physics must provide a mechanism to cut off these divergences,
without altering the phenomenological success of the SM. In order
to have a natural reductive effect, this in general requires
partner particles associated with a new symmetry beyond the SM.
Interestingly, these new particles are favored by the existence of
dark matter in the universe as the SM cannot provide a viable
candidate.

The most popular approach to solve such a naturalness/fine-tuning
problem is weak scale supersymmetry implemented with
$R$-parity,\cite{MSSM} where the standard particles share the SM
gauge symmetry with their supersymmetric partners. However a new
problem comes forth, namely, the mass of some partner particles in
the minimal supersymmetric extension of the SM implied by
radiative corrections to the Higgs mass falls below the energy
scale currently explored.\cite{PDG} As none of these partners has
been observed yet, the minimal version has to be expanded, for
instance, with extra fields as in the next-to-minimal
supersymmetric model.\cite{NMSSM1,NMSSM2}

Another approach contemplates an increase of the separation
between the Higgs mass and the above cutoff based on the idea that
the Higgs boson is light because it is in part a pseudo-Goldstone
boson of a broken global symmetry. This possibility has been
developed in the so-called little Higgs models with
$T$-parity,\cite{Cheng1,Cheng2} twin Higgs models with twin
symmetry,\cite{Chacko1}$^{\mbox{--}}$\cite{Twin} and mirror matter
models with mirror symmetry,\cite{Foot1,Foot2} where the discrete
symmetries relate the SM particles with their partners alleviating
the naturalness problem of the weak scale. Replication of the SM
particles is typical of these models beyond the SM.

While all these developments pursue a deep understanding of the
electroweak gauge symmetry breaking and mass generation issues,
they do not shed any light on the fermion family problem. These
are related questions which should be answered by the same type of
new physics beyond the SM. With so many families replicas, it is
conceivable that the symmetry associated with the new partner
particles and the symmetry associated with the existence of three
fermion generations be related to each other in a unified
description. In the end, they all are part of the same problem of
family reproductions.

In this paper we argue that a symmetry indeed exists which
accounts for the triplication of fermion families and the twin or
mirror partners proposed to deal with the naturalness problem of
the weak scale, then adding robustness to the twin and mirror
symmetries. This symmetry is presymmetry,\cite{EAM1,EAM2} a
predynamical symmetry hidden by the nontrivial topology of weak
gauge fields which addresses the question of quark--lepton
symmetry exhibited plainly in the electroweak sector of the SM
when right-handed neutrinos are included. In Sec.~2 we discuss the
main features of presymmetry and the rationale that sustains our
approach to the solution of the family problem. In Sec.~3 we
provide motivations to go beyond the SM with presymmetry,
emphasizing the duplication of the SM particles with twin or
mirror partners to have a residual presymmetry and so address the
naturalness problem. The conclusion is given in Sec.~4.

\section{Presymmetry}
\label{presym}

The quark--lepton symmetry has been extended from weak to
electromagnetic interactions via a mechanism of charge
fractionalization with topological attributes as in condensed
matter physics.\cite{EAM1,EAM2} The approach is based on the
charge symmetry between quarks and leptons. For an arbitrary weak
hypercharge $Y$ defined in terms of the electric charge $Q$ and
the third component of weak isospin $T_3$ according to
\begin{equation}
Q = T_{3} + a \; Y \, , \label{charge}
\end{equation}
the quark--lepton charge symmetry is recognized from the following
relation between hypercharges:\cite{EAM2}
\begin{eqnarray}
\begin{array}{l}
Y(q_{L,R}) = \displaystyle Y(\ell_{L,R}) - \frac{2}{3 a} \,
(3B-L)(\ell_{L,R}) \, , \\ [12pt] Y(\ell_{L,R}) = \displaystyle
Y(q_{L,R}) - \frac{2}{3 a} \, (3B-L)(q_{L,R}) \, ,
\end{array}
\label{chargesym}
\end{eqnarray}
where $q_{L,R}$ and $\ell_{L,R}$ refer to quark and lepton weak
partners within each of the three known families, right-handed
neutrinos of $Y=0$ included. Values of $a$ used in the literature
are $a = 1$ and $a = 1/2$. Although the value of the global part
depends upon the hypercharge normalization, the charge symmetry is
present anyway.

Presymmetry has to do with the exact correspondence between charge
values of quarks and leptons if the global part was not present,
clearly shown by Eq.~(\ref{chargesym}). We set forth this
connection in the electroweak quark--lepton symmetry principle:
there is an intrinsic underlying equality of quark and lepton
electroweak charges, regardless of how these charges are defined.
We note that Majorana neutrinos are excluded from this symmetric
picture of quarks and leptons which makes sense only for Dirac
neutrinos.

\subsection{Prequarks and preleptons}
\label{preQ}

To discern the charge symmetry and the charge dequantization
hidden in Eq.~(\ref{chargesym}), we introduce the new primary
states of prequarks and preleptons, denoted by $\hat{q}$ and
$\hat{\ell}$, which have the same quantum numbers of quarks and
leptons, respectively, except hypercharge values. Hypercharges of
prequarks and preleptons are the same as their lepton and quark
weak partners. Specifically, we follows Eq.~(\ref{chargesym}) to
have\cite{EAM2}
\begin{eqnarray}
\begin{array}{l}
Y(q_{L,R}) = \displaystyle Y(\hat{q}_{L,R}) - \frac{2}{3 a} \,
(B-3L)(\hat{q}_{L,R}) \, , \\ [12pt] Y(\ell_{L,R}) = \displaystyle
Y(\hat{\ell}_{L,R}) - \frac{2}{3 a} \, (B-3L)(\hat{\ell}_{L,R}) \,
,
\end{array}
\label{hathyper}
\end{eqnarray}
with prequark--lepton and prelepton--quark charge symmetries given
by
\begin{eqnarray}
\begin{array}{c}
 Y(\hat{q}_{L,R}) = Y(\ell_{L,R}) \, , \qquad Y(\hat{\ell}_{L,R})
= Y(q_{L,R}) \, , \\ [12pt] (B-3L)(\hat{q}_{L,R}) =
(3B-L)(\ell_{L,R}) \, , \qquad (B-3L)(\hat{\ell}_{L,R}) =
(3B-L)(q_{L,R}) \, ,
\end{array}
\label{hatql}
\end{eqnarray}
having $(B-3L)(\hat{\ell}_{L,R})= - (B-3L)(\hat{q}_{L,R})$. In
Eq.~(\ref{hathyper}) the combination $B-3L$ is in place of $3B-L$
because prequarks are the ones that now possess the lepton charges
and preleptons the quark charges, as indicated in
Eq.~(\ref{hatql}). From the latter we readily obtain
$B(\hat{q}_{L,R})=-1$ and $L(\hat{\ell}_{L,R})=-1/3$; these values
can be made positive if in Eqs.~(\ref{hathyper}) and (\ref{hatql})
we use $3L-B$ instead of $B-3L$. Thus, $B-L$ and    \linebreak
not $B-3L$ is the same for quarks and preleptons, as for prequarks
and leptons. We see that the $B-3L$ is essentially a bookkeeping
global charge based on counting such that three preleptons make a
system with one unit of $B-L$ charge, just as three quarks do.

We proceed with the charge symmetry exhibited in
Eq.~(\ref{chargesym}) and the charge dequantization described in
Eq.~(\ref{hathyper}) having $B$ and $L$ as ungauged global
symmetries, quarks and leptons as the ultimate constituents of
ordinary matter, and prequarks and preleptons as their basic bare
states. The global, robust against local interactions, piece of
hypercharge $2(B-3L)/3a$ gets a topological significance
associated with a topological charge or Pontryagin index which is
independent of the normalization used for hypercharge. This
implies a mixing of underlying local and topological charges, as
discussed in the following.

\subsection{Topological quarks and presymmetry}

We now implement topologically the hidden quark--lepton charge
symmetry shown above guided by the fact that any weak topological
feature cannot have observable effects at the zero-temperature
scale because of the smallness of the weak coupling. We then
introduce the principle of weak topological charge
confinement:\cite{EAM2} observable particles have no weak
topological charge, considering that the topological numbers are
carried by the vacuum in which particles exist. It is secondary to
that of gauge confinement, in the sense that electroweak forces by
themselves cannot lead to confinement of topologically nontrivial
particles. In the case of (pre)quarks, confinement is due to the
strong color force.

This principle guarantees that quarks and leptons are
topologically trivial and have no charge structure. Consequently,
the charge structure held by Eqs.~(\ref{hathyper}) and
(\ref{hatql}) does not apply to quarks, but to new entities that
we name topological quarks. The assignments of topological quarks
to the gauge groups of the SM, however, are the same of quarks.
There is electroweak symmetry between topological quarks and
preleptons, which may also be named topological leptons, as
between prequarks and leptons (see Eq.~(\ref{hatql})). Presymmetry
is the statement of this charge symmetry. Analytically, it is the
invariance of the bare electroweak Lagrangian under flavor
transformations of a $Z_2$ group which interchange topological
quarks (prequarks) and preleptons (leptons) weak partners, with no
change on gauge and Higgs fields.

The feature that in baryons quarks are confined in threes
containing each of the three colors requires at least a weak
topological charge associated with a bookkeeping $Z_3$ charge,
defining a nontrivial value $+1$ for topological quarks to have
the equivalence between three topological quarks and three quarks.
Hence, the~$3$ of this modulo charge in topological quarks, based
on counting, is due to the number of colors. By presymmetry,
preleptons also have a $Z_3$ charge equal to $+1$. We argue below
that the~$3$ of this modulo charge in preleptons is due to the
number of families. When the bookkeeping charge is $3$, the set
has no topological charge and trivial topology, although leptons
do not confine.

\subsection{Topological charge and charge normalization}
\label{topology}

It is assumed that prequarks and preleptons interact with the
standard gauge and Higgs fields through a Lagrangian like that of
the SM with quarks and leptons excepting hypercharge couplings and
incorporation of Dirac right-handed neutrinos. The cancellation of
gauge anomalies generated by their nonstandard charges leads to
the appearance of a topological charge. More specifically, in the
presymmetric scenario of prequarks and leptons, each prequark
changes its hypercharge by the same value, a charge shift which
can be written as follows:\cite{EAM2}
\begin{equation}
Y(\hat{q}) \rightarrow Y(\hat{q}) + \frac{n}{6a} \;
Q^{(3)}(\hat{q}) = Y(\hat{q}) - \frac{n}{6a} \; (B-3L)(\hat{q}) \,
, \label{norma}
\end{equation}
where $n$ is the topological charge of a $\mbox{SU(2)}_L$
instanton, with a $Z_{3}$ counting number $Q^{(3)}$ attached to it
equal to $\pm 1$ for nontrivial topology and $0$ for trivial
topology, just as if the topological charge were itself a $Z_{3}$
charge. It is due to the hypothesis of the approach. We define
$Q^{(3)}=-(B-3L)$ in accordance with Eq.~(\ref{hathyper}), noting
that three prequarks pass to a system of $Q^{(3)}=3 \, (=0)$, the
neutral element   \linebreak
of $Z_3$.

The required value for the topological index is $n=4$ because of
the gauge anomaly cancellation, demanded for gauge invariance and
renormalizability, and so consistency of the gauge theory. All
assignments of hypercharge (times the conventional parameter $a$)
of topological quarks are determined, resulting in the hypercharge
of quarks and their observed electric charges. The value singled
out for the topological charge does not depend on $a$, as
indicated in Subsec.~\ref{preQ}. Now, in order to get a lower
value one has to go beyond the SM. For instance, a topological
charge $n=2$ is obtained by a symmetric duplication of the
$\mbox{SU(2)}_L$ gauge group as in the $\mbox{SU(2)}_L \times
\mbox{SU(2)}_R \times \mbox{U(1)}_{B-L}$ left--right symmetric
models, which we will report on in a separate paper.

The first term of the right-hand side of Eq.~(\ref{norma}) is
associated with a local charge created by local fields and the
second with a topological charge related to a weak instanton which
cannot be generated by local operations. We then have here a
concrete example of a mechanism of charge fractionalization in
which states of local fields\,---\,prequarks\,---\,pass to states
with a topological character\,---\,topological quarks\,---\,which
allows local and topological charges to be mixed.\cite{Wilczek} It
is a hidden charge structure that explains the fractional charge
of topological quarks as in condensed matter physics.

It is concluded that topological quarks are involved in a vacuum
gauge field configuration of winding number $n_W = 4$, if we use
gauge freedom to set $n_W = 0$ for the one containing prequarks.
The transformation of prequarks into topological quarks is via an
Euclidean topological weak-instanton with topological charge
$n=4$, interpreted in Minkowski spacetime as a quantum-mechanical
tunneling event between vacuum states of weak $\mbox{SU(2)}_L$
gauge fields with different topological winding quantum numbers.
In this sense, prequarks and topological quarks reside in
different vacua. In other words, the difference between
topological quarks and prequarks is the nonequivalence between the
topological vacua of their weak gauge configurations, tunneled by
a weak four-instanton which carries the topological charge and
induces the universal fractional piece of charge needed for
normalization. Similarly, the passage from topological quarks to
quarks of same hypercharge values is via a $n=-4$ vacuum tunneling
weak-instanton, when three confined topological quarks of neutral
$Z_3$ total charge pass to three confined quarks of trivial
topology.

We remark that the charge normalization and the zero weak
topological charge in quarks and hadrons effectively remove the
extremely large time scale for the transitions from prequarks to
topological quarks and from the latter to quarks, associated with
the extreme smallness of the instanton transition probability.
Besides, these transitions do not happen in the real world
because, as argued in Sec.~\ref{residual}, prequarks and
topological quarks are not real dynamical entities.

Gauge anomaly cancellation and charge normalization in the
presymmetric scenario of topological quarks and preleptons are
done in a similar way.\cite{EAM2} The topological structure and
charge dequantization in preleptons, which are symmetric to the
ones in topological quarks, are annulled by the four-instanton
effect, leading to leptons with trivial topology and charge as in
prequarks (see Eqs.~(\ref{hathyper}) and (\ref{hatql})).

\subsection{Number of families}

At the level of quarks and leptons, the number~$3$ in
Eq.~(\ref{chargesym}), which is also the order of the additive
group $Z_{3}$ associated with the topological charge in the
framework of topological quarks, goes with the color number
because of the correlation in quarks between the baryon number and
the number of colors. Indeed, the occurrence of three colors
accounts for the fact that baryons made of three quarks of baryon
number $B=1/3$, have $B=1$; the three quarks containing each of
the three possible colors of color charge. This can be
accommodated in the relation $B = 1 / N_{c}$, in the case of
quarks. But, in the hidden scenario of prequarks and preleptons,
the~$3$ in Eq.~(\ref{hathyper}) cannot be conceived in the same
way by the facts that $B=-1$ for prequarks and that color is not a
prelepton quantum number. The number $3$ at the two levels of
descriptions given in Eqs.~(\ref{chargesym}) and (\ref{hathyper})
therefore has to be interpreted differently. At the hidden level,
it must be associated with a numerable property of prequarks and
preleptons and the number of families $N_{f}$ is the only other
available degree of freedom. Thus, whereas the partition of
topological charges in the scenario with topological quarks
depends on the number of colors, the partition in the scenario
with preleptons has to be in conformity with the number of
families; prequarks and leptons are topologically trivial. To
relate the~$3$ in Eq.~(\ref{hathyper}), which is the order of the
additive group $Z_3$ for the topological charge in colorless
preleptons, with the number of families becomes inevitable if we
assume that this number in $Z_3$ must be explained by physics of
the SM.\cite{EAM2} No other new physics is needed to understand
that number~$3$. It would be surprising to not have such a
connection, considering the fact that the SM offers no reason for
the triplication of     \linebreak
families.

In the end, preleptons have a lepton number marked by the number
of families: $L = - 1 / N_{f}$, as inferred from
Subsec.~\ref{preQ}. Thus, for presymmetric prequarks and leptons
one has $B-L=-1$, whereas for preleptons and topological quarks
presymmetry leads to $B-L=1/3$, with $N_{f}=N_{c}=3$. Now, the
prequark hypercharge shift in Eq.~(\ref{norma}) can be written as
$\Delta Y = 2 / aN_{f}$, displaying the expected meaning of the
$3$ in the hypercharge relationships of Eq.~(\ref{hathyper}).

The solution of the family problem is given by presymmetry which
demands the same number of families of quarks and leptons and the
exact correspondence between this number and that of quark colors.

\section{On a Residual Presymmetry}
\label {residual}

Electroweak presymmetry is hidden at the level of standard quarks
and leptons. Fractional charge is generated in a peculiar manner
but only mathematically. Neither topological quarks, prequarks and
preleptons are real dynamical entities with definite mass values
nor the associated presymmetry has a mass scale breaking. They are
not the particles that do the job with the physical gauge and
Higgs fields of the SM. All of them are bare prestates of quarks
and leptons which are seen as convenient mathematical entities out
of which the actual particle states are built up. It is meant as a
scheme that guesses at a new hidden charge symmetry, presymmetry,
which embraces quarks and leptons. If taken as a real dynamical
model, it possesses serious problems for presymmetric topological
quarks, prequarks and preleptons cannot be physical states; in
simple terms, these do not exist. For instance, transitions from
prequarks to topological quarks and from the latter to quarks
would be faced badly with the negligible smallness of the
instanton transition probability if the former were real objects,
but they are not. This is what allows to define the mixing of
local and topological charges in Eq.~(\ref{norma}).

In spite of that, the proposal provides a theoretical framework
which has many physical implications:\cite{EAM2} it explains the
fractional charge of quarks and the quark--lepton charge
relations; it states that the number of fermion generations has to
be equal to the number of quark colors; it predicts $B-L$
conservation and the Dirac character of massive neutrinos; it
accounts for the topological charge conservation in quantum flavor
dynamics; it explains charge quantization and the no observation
of fractionally charged hadrons.

Even so, there is nothing physically new and nothing has been
altered at the level of the SM. This is disturbing, because one
may expect some other new physics to account for the above
implications and therefore ask for the Occam's razor: ``Entities
should not be multiplied unnecessary.'' To avoid it, a residual
presymmetry in the sense of Ekstein\cite{Ekstein1,Ekstein2} has to
be generated.  Besides, it is really difficult to accept that the
hidden picture of the discrete presymmetry cannot be tested; if
this is the case, it is impossible to either verify or falsify the
proposal. These are strong motivations to take presymmetry beyond
the SM. Other reasons are to extend presymmetry from matter to
forces and from the electroweak to the strong sector, i.e. to have
presymmetry for the full Lagrangian of fundamental interactions,
then acquiring more significance with a strong influence on the
course of the new physics. A residual presymmetry based on these
motivations requires a doubling of the SM particles, whose
existence will make the substantiation of the proposition by
leading to new experimentally observable predictions. Due to the
connection between the number of fermion generations and the
number of quark colors, the new families must be nonsequential,
duplicating the gauge groups. On the other hand, due to its
topological character, presymmetry is unrelated to the energy
scale and appears to be transverse to everything, prompting in
particular its enlargement to the forces of the SM independently
of the ultraviolet completion of the theory.

\subsection{Twin symmetry from presymmetry}

The simplest duplication of the SM keeps spin and handness. It is
a plain copy of the SM particles much as the second and third
generations of quarks and leptons are mere copies of the first
generation; copies all that can be regarded as implied by
presymmetry. Now we describe how under this replication of
particles a residual presymmetry comes out and extends from matter
to forces and from weak to strong interactions.

On the one hand, it is the hidden charge symmetry relating quark
and lepton multiplets, as explained in Sec.~\ref{presym}, and
their respective partners denoted by tildes:
\begin{eqnarray}
\begin{array}{l}
(u_L,d_L) \leftrightarrow (\nu_L,e_L) , \qquad u_{R}
\leftrightarrow \nu_{R} , \qquad d_{R} \leftrightarrow e_{R} ,
\\ [12pt] (\tilde{u}_L,\tilde{d}_L)
\leftrightarrow (\tilde{\nu}_L,\tilde{e}_L) , \qquad \tilde{u}_{R}
\leftrightarrow \tilde{\nu}_{R} , \qquad \tilde{d}_{R}
\leftrightarrow \tilde{e}_{R} ,
\end{array}
\label{Presymm}
\end{eqnarray}
where right-handed neutrinos have been included. The underlying
presymmetry between fermions is hidden by the charge shifts
induced by the topological charges associated with the
configurations of weak gauge fields. Gauge and Higgs fields are
not changed by the presymmetric interchanges that leave invariant
the electroweak part of the bare Lagrangian.

On the other hand, there is a similar hidden charge symmetry
between quarks and the partners of leptons, and between their
copies, respectively:
\begin{eqnarray}
\begin{array}{l}
(u_L,d_L) \leftrightarrow (\tilde{\nu}_L,\tilde{e}_L) , \qquad
u_{R} \leftrightarrow \tilde{\nu}_{R} , \qquad d_{R}
\leftrightarrow \tilde{e}_{R} , \\ [12pt]
(\tilde{u}_L,\tilde{d}_L) \leftrightarrow (\nu_L,e_L) , \qquad
\tilde{u}_{R} \leftrightarrow \nu_{R} , \qquad \tilde{d}_{R}
\leftrightarrow e_{R} .
\end{array}
\label{partnerPresymm}
\end{eqnarray}
Here the electroweak symmetry which interchanges the gauge and
Higgs bosons with their partners requires that the corresponding
coupling constants be equal.

The $Z_2$ symmetries of Eqs.~(\ref{Presymm}) and
(\ref{partnerPresymm}) lead to the following one between quarks
and their partners, and between leptons and their duplicates:
\begin{eqnarray}
\begin{array}{c}
(u_L,d_L) \leftrightarrow (\tilde{u}_L,\tilde{d}_L) , \qquad u_{R}
\leftrightarrow \tilde{u}_{R} , \qquad d_{R} \leftrightarrow
\tilde{d}_{R} , \\ [12pt] (\nu_L,e_L) \leftrightarrow
(\tilde{\nu}_L,\tilde{e}_L) , \qquad \nu_{R} \leftrightarrow
\tilde{\nu}_{R} , \qquad e_{R} \leftrightarrow \tilde{e}_{R} .
\end{array}
\label{resPresymm}
\end{eqnarray}

Besides, there is symmetry between electroweak gauge and Higgs
bosons and their partners, with same coupling constants. This
$Z_2$ symmetry, but not the others, remains exact after the
underlying charge normalization on fermions. Moreover, it extends
to strong interactions for equal gauge couplings of the two color
groups. In this case, an observable residual $Z_2$ symmetry
exists, the required residual presymmetry, which includes the
strong sector, relates every particle of the SM with its partner
particle and constrains the corresponding coupling constants to be
equal, just as in twin matter models with twin symmetry. As a
consequence, the existence of two symmetric Higgs doublets
alleviates the hierarchy
problem.\cite{Chacko1}$^{\mbox{--}}$\cite{Twin} This is discussed
in the following.

Under a full duplication of the SM, there are two renormalizable
couplings between particles of the SM and their partners allowed
by gauge invariance: $\lambda \phi ^{\dagger} \phi
\tilde{\phi}^{\dagger} \tilde{\phi}$ and $\epsilon B^{\mu\nu}
\tilde{B}_{\mu\nu}$, where $\phi$, $\tilde{\phi}$ are the Higgs
doublets of the SM and its copy respectively and $B^{\mu\nu}$,
$\tilde{B}^{\mu\nu}$ are the hypercharge field strength tensors.
In reference to the Higgs sector, there is a limit in which the
Higgs scalar may be treated as a pseudo-Goldstone boson. To see
it, the Higgs potential is written as
\begin{eqnarray}
V & = & - \mu^2 ( \phi^{\dagger} \phi + \tilde{\phi}^{\dagger}
\tilde{\phi} ) + \lambda ( \phi^{\dagger} \phi +
\tilde{\phi}^{\dagger} \tilde{\phi} )^{2} + \delta [
(\phi^{\dagger} \phi)^{2} + (\tilde{\phi}^{\dagger}
\tilde{\phi})^{2} ] , \label{potential}
\end{eqnarray}
where the term proportional to $\lambda$ contains the above
coupling of Higgs-doublet partners. The potential maintains a
$\mbox{U}(4)$ global symmetry in the limit $\delta \rightarrow 0$.
The model presents two nontrivial vacua which rely on whether
$\delta > 0$ (symmetric vacuum) or $\delta < 0$ (asymmetric
vacuum).\cite{Twin} The symmetric vacuum, where both Higgs
doublets get the same vacuum expectation values, is the
interesting case. Here, $\langle \phi \rangle = \langle
\tilde{\phi} \rangle = v$ with $v^2 = \mu^2 / (4\lambda+2\delta)$.
Although gauge and Yukawa couplings violate the global symmetry,
the discrete symmetry that interchanges particles and partners is
respected.

In the SM, the most significant quadratically divergent one-loop
contributions to the Higgs potential involve the top quark, the
gauge bosons, and the Higgs scalar. Keeping just the one-loop top
quark correction,
\begin{equation}
\mu^{2} = \mu_{\circ}^{2} + a_{t} \Lambda_{t}^{2} ,
\end{equation}
where $\mu_{\circ}$ is the bare parameter, $a_{t}=3\lambda_{t}^{2}
/ 8\pi^{2}$, $\lambda_{t}=m_{t} / v_{t} \sim 1$ is the top quark
Yukawa coupling constant, and $\Lambda_{t}$ is the cutoff from new
physics. In the extended model with $Z_2$ symmetric partners, the
quadratic divergence maintains its form in both sectors and so the
$\mbox{U}(4)$ symmetry. The spontaneous symmetry breaking
$\mbox{U}(4) \rightarrow \mbox{U}(3)$ in the limit $\delta
\rightarrow 0$ leads to one massless Higgs boson, as expected.
Corrections to the Higgs quartic interactions that are not
invariant under the global symmetry, such as the $\delta$ term in
Eq.~(\ref{potential}), provide mass to the Higgs of order the weak
scale.\cite{Chacko1,Chacko2}

The quadratic divergence in $\mu^2$ from the top quarks is
alleviated in the duplicated model. In fact, a measure of
fine-tuning is
\begin{equation}
\displaystyle \left( \frac{\delta \mu^{2}}{\mu^{2}} \right)_{t} =
\frac{a_t \Lambda_t^2}{\mu^2} ,
\end{equation}
with $\mu^{2} = m_{h}^{2} / 2 = 2 v^2 \lambda$ in the SM and
$\mu^{2} = m_{+}^{2} / 2$ in the duplicated model, where
$m_{+}^{2} = 4 v^2 (2 \lambda + \delta)$ is the mass of the
heavier physical Higgs boson and $m_{-}^{2} = 4 v^2 \delta$ is
that of the lighter. In the SM, the bound from precision
electroweak measurements is $m_h < m_{EW} \approx 186$
GeV.\cite{PDG} In the duplicated model, the bound is $m_{+} m_{-}
< m_{EW}^{2}$. Thus, because of a large $m_{+}$, the fine-tuning
in the $\mu^2$ parameter due to the top quark is alleviated.

More quantitatively, the scale of new physics depends on the
amount of fine-tuning that is allowed. A model is considered ideal
if $|\delta\mu^2/\mu^2| \lesssim 5$, corresponding to no
significant electroweak fine-tuning.\cite{Gunion} At the
experimental limit of about 114 GeV for $m_h$ in the SM and
$m_{-}$ in the duplicated model, and taking $m_{+}$ to the largest
value consistent with electroweak precision tests, the ideal value
of the upper limit on $\Lambda_t$ is
\begin{equation}
\Lambda_t = \frac{2\pi}{\sqrt{3}\lambda_t} \; m_h \left|
\frac{\delta \mu^2}{\mu^2} \right|^{1/2}_{t} \sim 0.9 \;
\mbox{TeV}
\end{equation}
in the SM and
\begin{equation}
\Lambda_t = \frac{2\pi}{\sqrt{3}\lambda_t} \; m_{+} \left|
\frac{\delta \mu^2}{\mu^2} \right|^{1/2}_{t} \sim 2.5 \;
\mbox{TeV}
\end{equation}
in the duplicated model, which shows the improvement of
naturalness of the Higgs sector. This cutoff can be scaled up by
allowing a moderate fine-tuning.

Regarding the one-loop Higgs correction to the quadratic
divergences in the SM, we have
\begin{equation}
\mu^{2} = \mu_{\circ}^{2} - a_{H} \Lambda_{H}^{2} ,
\end{equation}
where $a_{H}=3\lambda / 8\pi^{2}$ with $\lambda$ being the quartic
coupling constant of the Higgs potential and $\Lambda_H$ the
cutoff from new physics for this divergence. It leads to the
result
\begin{equation}
\displaystyle \left( \frac{\delta \mu^{2}}{\mu^{2}} \right)_H = -
\, \frac{a_H \Lambda_H^2}{\mu^2} .
\end{equation}
In the duplicated model with $Z_2$ symmetry the correction goes
with $a_{H}=(5\lambda+3\delta) / 8\pi^{2}$. The values for the
upper bound on the cutoff are
\begin{equation}
\Lambda_H = \frac{4\pi}{\sqrt{3}} \; v \left| \frac{\delta
\mu^2}{\mu^2} \right|^{1/2}_{H} \sim 2.8 \; \mbox{TeV}
\end{equation}
in the SM and
\begin{equation}
\Lambda_H = \frac{4\pi\sqrt{2}}{\sqrt{5+\gamma}} \; v \left|
\frac{\delta \mu^2}{\mu^2} \right|^{1/2}_{H} \sim 3.0 \;
\mbox{TeV}
\end{equation}
with $\gamma=m_{-}^2 / m_{+}^2$ in the extended model. Here only a
little increase in the scale of new physics relative to the SM is
feasible. In the case of gauge boson loops, contributions to
quadratic divergences have the same form as from Higgs bosons and
also become smaller in magnitude compared with that from top
quark.

Thus, $\Lambda_t$ sets the scale of new physics in this domain of
parameters where a relatively light Higgs, as favored by precision
electroweak data, is assumed. Since $\Lambda_t \sim 0.9$ TeV of
the SM is within reach of LHC, manifestations of the new physics
are expected. In the case of the duplicated model, if $\Lambda_t
\sim 2.5$ TeV were outside reach of LHC, this extended model with
no other new physics would be perfectly natural. There would be
consistency with any bound from precision electroweak measurements
because partner particles are neutral with respect to the SM gauge
interactions. If the Higgs boson does not exists and other
symmetry breaking mechanism is operative, the duplication of the
SM goes anyway.

\subsection{Mirror symmetry from presymmetry}

There is an alternative, also simple copy of the SM leading to a
residual presymmetry. It is the mirror-symmetric case where an
exact parity symmetry is claimed as due. Left-handed weak gauge
bosons act on SM particles and right-handed ones on their
partners. Now, instead of Eq.~(\ref{Presymm}), the hidden $Z_2$
symmetry is according to
\begin{eqnarray}
\begin{array}{l}
(u_L,d_L) \leftrightarrow (\nu_L,e_L) , \qquad u_{R}
\leftrightarrow \nu_{R} , \qquad d_{R} \leftrightarrow e_{R} ,
\\ [12pt] (\tilde{u}_R,\tilde{d}_R)
\leftrightarrow (\tilde{\nu}_R,\tilde{e}_R) , \qquad \tilde{u}_{L}
\leftrightarrow \tilde{\nu}_{L} , \qquad \tilde{d}_{L}
\leftrightarrow \tilde{e}_{L} .
\end{array}
\end{eqnarray}
In place of the charge symmetry in Eq. (\ref{partnerPresymm}), we
have
\begin{eqnarray}
\begin{array}{l}
(u_L,d_L) \leftrightarrow (\tilde{\nu}_R,\tilde{e}_R) , \qquad
u_{R} \leftrightarrow \tilde{\nu}_{L} , \qquad d_{R}
\leftrightarrow \tilde{e}_{L} , \\ [12pt]
(\tilde{u}_R,\tilde{d}_R) \leftrightarrow (\nu_L,e_L) , \qquad
\tilde{u}_{L} \leftrightarrow \nu_{R} , \qquad \tilde{d}_{L}
\leftrightarrow e_{R} .
\end{array}
\end{eqnarray}
Finally, instead of Eq. (\ref{resPresymm}), we obtain mirror
symmetry
\begin{eqnarray}
\begin{array}{c}
(u_L,d_L) \leftrightarrow (\tilde{u}_R,\tilde{d}_R) , \qquad u_{R}
\leftrightarrow \tilde{u}_{L} , \qquad d_{R} \leftrightarrow
\tilde{d}_{L} , \\ [12pt] (\nu_L,e_L) \leftrightarrow
(\tilde{\nu}_R,\tilde{e}_R) , \qquad \nu_{R} \leftrightarrow
\tilde{\nu}_{L} , \qquad e_{R} \leftrightarrow \tilde{e}_{L} .
\end{array}
\end{eqnarray}
All gauge coupling constants in both sectors are also related by
mirror parity. The Higgs sector is as in the above model,
alleviating the hierarchy problem in the same way, as in mirror
matter models with mirror symmetry.\cite{Foot1} Thus, the residual
presymmetry that is demanded can be seen as being the cause of
twin symmetry or mirror symmetry. It is worth noting that these
symmetries emerge independently of the solution of the quadratic
divergence problem.

Presymmetry remains hidden and the model is in trouble if there is
no copy of the SM particles. Majorana neutrinos and sequential
families, such as a fourth generation, also bring problems to the
idea of presymmetry which requires equal numbers of fermion
families and quark colors. If anything of this could occur, one
would go back to the starting point of the model and state that
the quark--lepton charge symmetry which supports presymmetry is
just an accidental interplay of quantum numbers, which is really
hard to be accepted.

\section{Conclusion}

Presymmetry is a hidden electroweak symmetry of the SM that
embraces quarks and leptons with many physical implications. It
explains the fractional charge of quarks and the quark--lepton
charge relations. It states that the number of fermion generations
has to be equal to the number of quark colors. It predicts $B-L$
conservation and the Dirac character of massive neutrinos. It
accounts for the topological charge conservation in quantum flavor
dynamics. It explains charge quantization and the no observation
of fractionally charged hadrons.

In order to be substantiated, however, it must be extended beyond
the SM                             \linebreak
as to have a residual presymmetry that includes partner particles
and the strong sector. And the case has been made in which a close
relation exists between this residual presymmetry and the twin and
mirror symmetries proposed in the literature to mitigate the
naturalness problem of the weak scale. The duplication of fermion
families and gauge bosons of the SM is predicted by presymmetry
independently of the ultraviolet completion of the theory, even if
the Higgs particle is discarded by experiment, which adds
robustness to twin and mirror symmetries. The established
connection is so strong that the search for twin or mirror
symmetry becomes the possible test of presymmetry. Experimentally
observable predictions are extracted from twin or mirror matter
models.\cite{Twin,Ignatiev,Schabinger} If the physics beyond the
SM repairs its left--right asymmetry, mirror symmetry should be
the one realized in     \linebreak    nature.

\section*{Acknowledgments}

We would like to thank J. Gamboa for helpful comments. This work
was supported by the Departamento de Investigaciones
Cient\'{\i}ficas y Tecnol\'ogicas, Universidad de Santiago de
Chile, Usach.


\vspace{2pt}

\begin{thebibliography}{99}
\bibitem{PDG} Particle Data Group (K. Nakamura \emph{et al.}),
{\it J. Phys. G} {\bf 37}, 075021 (2010).

\bibitem{Barbieri} R. Barbieri and A. Strumia, arXiv:hep-ph/0007265.

\bibitem{MSSM} D. J. H. Chung \emph{et al.}, {\it Phys. Rep.} {\bf 407},
1 (2005).

\bibitem{NMSSM1} M. Bastero-Gil \emph{et al.}, {\it Phys. Lett. B} {\bf
489}, 359 (2000).

\bibitem{NMSSM2} U. Ellwanger, C. Hugonie and A. M. Teixeira,
arXiv:0910.1785.

\bibitem{Cheng1} H. C. Cheng and I. Low, {\it J. High Energy Phys.}
{\bf 0309}, 051 (2003).

\bibitem{Cheng2} H. C. Cheng and I. Low, {\it J. High Energy Phys.}
{\bf 0408}, 061 (2004).

\bibitem{Chacko1} Z. Chacko, H. S. Goh and R. Harnik, {\it Phys. Rev.
Lett.} {\bf 96}, 231802 (2006).

\bibitem{Chacko2} Z. Chacko, Y. Nomura, M. Papucci and G. Perez,
{\it J. High Energy Phys.} {\bf 0601}, 126 (2006).

\bibitem{Twin} R. Barbieri, T. Gregoire and L. J. Hall,
arXiv:hep-ph/0509242.

\bibitem{Foot1} R. Foot and R. R. Volkas, {\it Phys. Lett. B} {\bf 645},
75 (2007).

\bibitem{Foot2} R. Foot, H. Lew and R. R. Volkas, {\it Phys. Lett. B}
{\bf 272}, 67 (1991).

\bibitem{EAM1} E. A. Matute, {\it Phys. Lett. B} {\bf 538}, 66 (2002).

\bibitem{EAM2} E. A. Matute, {\it Int. J. Mod. Phys. A} {\bf 22},
3669 (2007).

\bibitem{Wilczek} F. Wilczek, arXiv:cond-mat/0206122.

\bibitem{Ekstein1} H. Ekstein, {\it Phys. Rev.} {\bf 153}, 1397 (1967).

\bibitem{Ekstein2} H. Ekstein, {\it Phys. Rev.} {\bf 184}, 1315 (1969).

\bibitem{Gunion} J. F. Gunion, arXiv:0804.4460.

\bibitem{Ignatiev} A. Y. Ignatiev and R. R. Volkas, {\it Phys. Lett.
B} {\bf 487}, 294 (2000).

\bibitem{Schabinger} R. Schabinger and J. D. Wells, {\it Phys. Rev. D}
{\bf 72}, 093007 (2005).
\end{thebibliography}
\end{document}